\documentclass[aps,pra,twocolumn,superscriptaddress,showpacs,longbibliography]{revtex4-1}
\usepackage{amssymb}
\usepackage{hyperref}
\usepackage{bm}
\usepackage{natbib}
\usepackage{graphicx}
\usepackage{psfrag}
\usepackage{subfigure}
\usepackage{verbatim}
\usepackage{color}
\usepackage{amsmath,amssymb}

\newcommand{\bea}{\begin{eqnarray}}
\newcommand{\eea}{\end{eqnarray}}

\begin{document}
\title{Fracture strength: Stress concentration, extreme value statistics and the
fate of the Weibull distribution}
\author{Zsolt Bertalan}
\affiliation{ISI Foundation, Via Alassio 11/c, 10126 Torino, Italy.}
\author{Ashivni Shekhawat}
\affiliation{Department of Materials Science and Engineering, University of California, Berkeley, California 94720, United States}
\affiliation{Materials Science Division, Lawrence Berkeley National Laboratory, Berkeley, California 94720, United States}
\author{James P. Sethna}
\affiliation{LASSP, Physics Department, Cornell University, Ithaca, NY 14853-2501, United States}
\author{Stefano Zapperi}
\affiliation{CNR, IENI, Via Roberto Cozzi 53, 20125 Milano, Italy.}
\affiliation{ISI Foundation, Via Alassio 11/c, 10126 Torino, Italy.}

\date{\today}


\begin{abstract}

The statistical properties of fracture strength of brittle and quasi-brittle materials are often described in terms of the Weibull distribution. The weakest-link hypothesis, commonly used to justify it, is however expected to fail when fracture occurs after significant damage accumulation. Here we show that this implies that the Weibull distribution is unstable in a renormalization group sense for a large class of quasi-brittle materials. Our theoretical arguments are supported by numerical simulations of disordered fuse networks. 
We also find that for brittle materials such as ceramics, the common assumption that the 
strength distribution can be derived from the distribution of pre-existing micro-cracks 
by using Griffith's criteria is invalid. We attribute this discrepancy to crack bridging. 
Our findings raise questions about the applicability of Weibull statistics to most practical cases.
\end{abstract}

\maketitle


\section{Introduction}
The applicability of the Weibull distribution to describe the fracture strength of brittle and quasi-brittle 
materials has been a topic of intense debate~\cite{danzer2007,dieter2002,Doremus1983,Jayatilaka1983,Manzato2012,nohut2012,bazant1991,rozenblat2011,basu2009}. 
Several experimental studies argue that the Weibull distribution is not always the best statistical distribution to 
fit fracture data~\cite{nohut2012,danzer2007,dieter2002,rozenblat2011,hosson1991,alava1998,baxevanakis1993} (numerous other argue otherwise), 
particularly for quasi-brittle materials that have significant precursor damage. 
These observations demand a general theoretical explanation. 
The suggested explanations for these empirical observation includes bimodal or multimodal 
flaw size distribution~\cite{danzer2007,jakus1981,orlovskaja2000,peterlik2001}, R-curve behavior~\cite{nohut2012}, small size of the datasets~\cite{danzer2007,basu2009}, 
and thermal activated crack nucleation \cite{santucci2007,vanel2009}. 
Here we provide a general explanation for these observations
by showing that the Weibull distribution is unstable in the renormalization
group sense for quasi-brittle materials, and thus not applicable at long length scales.
\par
In deriving the Weibull distribution of fracture strengths it is invariably assumed 
that the material volume has a population of non-interacting crack-like defects,
and fracture happens as soon as the weakest of these defects starts to grow~\cite{weibull,Heinisch1978,Sutcu1989}.
This assumption is also known as the `weakest-link hypothesis'. 
Experimental observations suggest that this assumption does not hold for a large class of quasi-brittle materials.
These materials, including paper~\cite{Alava2002}, granite~\cite{lockner1991quasi,garcimartin1997}, 
antler bone~\cite{Zioupos1994203}, wood~\cite{garcimartin1997,Tschegg2000}, and composites~\cite{yukalov2004,sornette1995} etc., typically ``crackle''~\cite{sethna2001,Manzato2012}, 
suggesting that several local cracks grow and get arrested prior to global fracture. 
Advanced composites are designed to fail gracefully, that is, they have multiple failures before the ultimate fracture. 
It is clear that for such materials the weakest defect does not dominate the fracture properties of the material, 
and the defects interact via elastic fields. 
The emergent scale-invariant properties of these interactions 
have been a topic of intense study in the statistical physics community~\cite{zapperi1997,hansen2003,shekhawat2013,alava2006}. 
Several researchers have used the Weibull theory to model these quasi-brittle materials. 
We show that even if the microscopic strength distribution 
is Weibull, the emergent distribution is significantly distorted due to elastic interactions
and metastability. Thus, the Weibull distribution is not stable in the renormalization group 
sense. We provide numerical evidence to support our theoretical claims. 
\par
For brittle materials such as glasses or ceramics that fracture catastrophically without precursor damage,  it is assumed that the distribution of fracture strength
can be derived from the distribution of flaw sizes by using Griffith's criteria (or 
equivalently the stress intensity approach) and ignoring effects such as 
crack bridging or coalescence~\cite{Jayatilaka1977,danzer2007}. 
For exponentially distributed cracks the fracture strength is expected to be described
by the Duxbury-Leath-Beale (DLB) distribution \cite{DLB,BealeDuxbury1988,Manzato2012},
while only in the case of power law distributed cracks one expects to obtain
the Weibull distribution \cite{Freudenthal1968}.
It is, however, challenging to measure the flaw size distribution experimentally,
and thus these assumptions are rarely verified empirically~\cite{chao1992,orlovskaja2000,zhang1998}.
One of the aims of this paper is use numerical simulations 
to show that the simple relations that are widely 
used in the literature are not accurate, and further study is needed to understand 
the discrepancy. This observation has important implications for material engineers 
who aim to improve the fracture properties of brittle materials by controlling the
micro-structure.
\par
In light of the above discussed limitations of the Weibull theory, what distribution should one use to 
fit fracture data? To answer this question, we consider 
two classes of fuse networks to model brittle and quasi-brittle materials. Both of the these 
models are derived from the classical fuse network models~\cite{DLB,RFM}. In the model for brittle
materials, the fuse network is seeded with power law distributed cracks with varying morphology. 
This is different from the classical diluted fuse network model which has an exponential distribution 
of cracks~\cite{DLB,BealeDuxbury1988,Manzato2012}. The model for quasi-brittle materials has a continuous distribution of fuse strengths, 
where each fuse strength is a random number drawn from a standard Weibull distribution. In this manner 
we can ensure that the microscopic strength distribution is Weibull, and study the emergent macroscopic 
distribution. This model differs from its counterparts in the literature~\cite{zapperi1997,Kahng1988} by the choice of 
the microscopic disorder, and enables a numerical study of the stability of the Weibull distribution.
Analyzing the simulations, we find that the recently proposed T-method provides a suitable alternative to fit the numerical data \cite{shekhawat2014}.  The method is general enough that it can be applied in a wide variety of cases.

\par

The rest of the paper is organized as follows. Section~\ref{sec:WeibullTheory} presents the basics of the classical  Weibull theory and the commonly used relation between the strength distribution and the defect size distribution.
The details of the numerical models used in this study are discussed in section~\ref{sec:RFM}. 
Section~\ref{sec:WeibStab} presents theoretical and numerical evidence to show that 
the Weibull distribution is unstable under coarse graining for quasi-brittle materials.
In section~\ref{sec:Mod} we present the numerical evidence to show that the relation between
the strength distribution and the flaw size distribution is nontrivial, and cannot be 
obtained by a straightforward application of the Griffith's criteria. We discuss the 
possible sources of the observed discrepancy. Section~\ref{sec:TMethod} presents
a comparison of the performance of the Weibull distribution and the recently proposed
T-method for fitting the simulation data for quasi-brittle fuse networks. The 
conclusions are presented in section~\ref{sec:conc}.

\section{Weibull Theory}
\label{sec:WeibullTheory}
In this section we review the classical Weibull theory in order to facilitate the discussion in
the following sections. 
We consider a material volume $V$ subjected to a stress field $\sigma(r)$. The material 
is supposed to have a density of defects of various shapes and sizes, such that $e^{-f(\sigma)}$ is the  probability of not finding a defect with critical stress less than $\sigma$ in a volume $V_0$ of the material. Here we assume that the stress in uniaxial and tensile; the case of full tensorial stress
is similar and is not presented here to avoid unnecessary notational complexity.
The volume $V_0$ is supposed to be large enough that it contains sufficient number of cracks, 
and yet small enough that the stress can be considered roughly constant across it; it is sometimes 
also called the representative volume element. 
$f(\sigma)$ is supposed to be a homogeneous material property. Then, the probability that 
the material volume $V$ will survive the stress field $\sigma(r)$ is given by 
\begin{equation}
S_V(\sigma) = \exp\left(  -\frac{1}{V_0}\int_V f(\sigma(r)) d{r}^3 \right).
\end{equation}
Weibull recognized that taking $f(\sigma) = (\sigma/\bar\sigma)^k$, where $\bar\sigma$ is a material dependent scale parameter, and $k$ is the material depended Weibull modulus, gave a good fit for several brittle materials, and introduced what is now known as the standard Weibull distribution \cite{weibull}
\begin{equation}
S_V(\sigma) = \exp\left(  -\frac{1}{V_0}\int_V \left( \frac{\sigma}{\bar\sigma}\right)^k d{r}^3 \right).
\label{eq:Weibull}
\end{equation}
It turns out that the empirical choice made by Weibull can be justified by a renormalization group 
calculation in which one writes recursive equations describing the failure distribution
as the scale is changed \cite{gyorgyi2010}. The Weibull distribution is one of the possible fixed points of the renormalization group transformation \cite{gyorgyi2010}.

The Weibull distribution can alternatively be derived by connecting the function $f(\sigma)$  to the microscopic defect size distribution. The basic calculation outlined in the remainder 
of this section can be found in a number of important references~\cite{Jayatilaka1977,danzer2007}.
According to Griffith's 
criteria, a crack of length $w$ is stable at applied normal stress $\sigma$ if 
\begin{equation}
K = \sigma Y w^{1/2} \leq K_{Ic},
\label{eq:StressIntensity}
\end{equation}
where $Y$ is the geometry factor of the crack, and $K_{Ic}$, the critical stress intensity factor, 
is a material property. The exponent of $1/2$ is applicable for ideally sharp cracks, and can have a 
different value for wedge shaped or blunted cracks. Thus, if we take $e^{-h(w)}$ to be probability 
that the volume element $V_0$ does not contain any crack longer than $w$, then we have 
\begin{equation}
f(\sigma) = h(K_{Ic}^2/\sigma^2Y^2).
\label{eq:DefectDensity}
\end{equation}
If the defect size (crack length) distribution is a power law with exponent $\gamma$, then $h(w) \sim w^{-\gamma}$, which
gives $f(\sigma) \sim (\sigma/\bar \sigma)^{2\gamma}$, where $\bar\sigma = K_{Ic}/Y$. Thus, a power law defect size 
distribution with exponent $\gamma$ leads to Weibull distribution of fracture strength with modulus $k = 2\gamma$. 
\par
As pointed out before, this entire analysis assumes 
that the flaws do not interact, and that the failure of the weakest flaw leads to the failure of the entire 
material volume. 
We also show in section~\ref{sec:WeibStab} relaxing the 
assumption that the weakest flaw leads to global failure has important consequences
and results in the strength distribution flowing away 
from the Weibull form.
Our numerical calculation reported in section~\ref{sec:Mod} sections show that crack bridging is an important form of 
crack interaction that can significantly alter the resulting Weibull modulus away from the
dilute limit (but does not change the Weibull form for power law distributed cracks). 


\section{The random fuse model}
\label{sec:RFM}
In this section we describe the computational model that we use for various classes 
of brittle and quasi-brittle materials. The theoretical arguments presented in
later sections benefit from having a concrete model as a point of reference.
We study several variants of the basic two dimensional random fuse model (RFM) \cite{RFM,DLB}. 
The RFM is a well accepted model of brittle fracture where each fuse represents a coarse grained 
material region (analogue of the classical representative volume element). 
The model consists of a set of conducting fuses  with unit conductivity $g_j=1$ and breaking
threshold $\sigma_j$, arranged on a 45$^\circ$-tilted square lattice composed by $L\times L$ 
nodes.  A unit voltage drop is applied along two parallel edges of the lattice while periodic boundary  conditions are imposed along the other two edges.  
The Kirchhoff equations are solved numerically using the algorithm proposed in Ref. \cite{nukalajpamg1} 
to determine the current flowing in each of the fuses.
We then evaluate the ratio between the current $i_j$ and the breaking threshold $\sigma_j$ and the fuse having the largest
value, $\mbox{max}_j \frac{i_j}{\sigma_j}$, is irreversibly removed (burnt).
The current is redistributed instantaneously after a fuse is burnt.
Each time a fuse is burnt, it is necessary to re-calculate the current distribution in the lattice.
The process of burning fuses, one at a time, is repeated until the lattice
system fails completely (becomes non-conductive).
The random fuse model is equivalent to a scalar elastic problem where we consider a pure anti-plane shear deformation. 
In this condition, the shear stress $\sigma$ is related to the total current $I$ by $\sigma = I/L$,
the shear strain $\epsilon$ to the voltage drop $v$ by $\epsilon=v/L$ and the conductivity
$g$ is equivalent to the shear modulus.
From the breaking sequence we can derive the current-voltage (or stress-strain) curve of the network under adiabatic loading
as discussed in Ref.~\cite{ANZrev2006}.

In this study we employ two different disorder distributions to model quasi-brittle and brittle materials:
\begin{itemize}
\item[i] {\it Weibull disorder (Quasi-brittle).} 
The fuse strength threshold is chosen to be  a random variable drawn from a Weibull distribution with modulus $k$, thus the survival probability of 
a fuse at applied stress $\sigma$ is $S_1(\sigma) = e^{-\sigma^k}$. Fuse networks with continuously 
distributed strengths have been studied previously~\cite{zapperi1997}. In those studies the thresholds were drawn from 
the uniform~\cite{zapperi1997},  power law~\cite{hansen2003,shekhawat2013}, and hyperbolic distributions~\cite{hansen2012}. However, the focus of those studies was 
on the morphology and dynamic properties, while we focus on strength. Further, by letting 
the local thresholds be Weibull distributed, we can directly study the stability of the Weibull
distribution under coarse graining.

\item[ii] {\it Diluted cracks (Brittle).} 
We remove a fraction $p$ of the fuses and assign the same breaking threshold (= 1)
to the intact fuses~\cite{DLB,Manzato2012}. 
We take $ 0.05 \leq p \leq 0.2$, thus keeping the initial damage fairly dilute
in order to avoid the phenomena that happens near the percolation threshold (at $p = 0.5$ for the tilted square lattice we are using). 
Note that the missing fuses are not chosen randomly, but rather in a way that they form a set of cracks with power 
law distributed crack lengths with $2.5\leq \gamma \leq 9$, where $\gamma$ is the 
exponent of the power law. 
We employ both straight and fractal flaws, grown by using self-avoiding random walks.
Fuse networks with diluted cracks were originally studied in Refs.~\cite{DLB,RFM}. However, in those studies 
the cracks lengths had an exponential distribution (as opposed to power law). Exponential 
distribution of defect sizes leads to a Gumbel type distribution of strengths, and thus 
are markedly different from our model.
\end{itemize}

For each case, we do extensive statistical sampling for network sizes $L = 32,\ 64,\ 128,\ 256,\ 512$.

\section{Stability of Weibull distribution for quasi-brittle materials}
\label{sec:WeibStab}

The standard Weibull distribution reported in  Eq.~\ref{eq:Weibull}  is derived under the assumption that the failure of the weakest flaw (or representative volume element)  leads to complete global failure. Under this assumption, if the strength distribution of the representative element is standard Weibull with modulus $k$, i.e.~$S_{V_0}(\sigma) = e^{-\sigma^k}$,
then the survival probability of the material volume $V$ is given by 
\begin{equation}
S_V(\sigma) = S_{V_0}(\sigma)^{V/V_0} = \exp\left( - \frac{V}{V_0} \sigma^k \right).
\label{eq:SimWeibull}
\end{equation}
In mathematical terms, we can say that the Weibull distribution is stable under
coarse graining: A system composed by subsystems described by the Weibull distribution is
itself described by the Weibull distribution. 

As we mentioned earlier, however, the weakest-link assumption is not accurate in quasi-brittle materials. We can then derive the condition for Eq. \ref{eq:SimWeibull} to remain valid if this assumption is relaxed. The stress at which the weakest flaw fails scales as
$\sigma_{\mathrm{min}} \sim (V_0/V)^k$. The failure of this volume element
enhances the stress on its neighbors due to stress concentration. However, 
the neighbors of the weakest flaw are typically not very weak, and we safely assume that
their strength is near the mean strength $\langle \sigma \rangle = 1$. 
Assuming that the stress concentration factor scales as $Y V_0^\beta$,
where $Y$ is a geometry factor, the neighboring 
volume element fails if $\sigma_{\mathrm{min}}YV_0^\beta > \langle \sigma \rangle$ which yields
$k \gtrsim \log V/V_0$ as the approximate condition for the validity of the weakest link 
hypothesis. Outside of this range the local failure of the weakest link does not trigger global failure. In this calculation we have ignored details and made several simplifications, 
thus it only gets the correct scaling.
\par
The above arguments show that the weakest link 
hypothesis is self-consistent, and the Weibull distribution is stable under coarse-graining only if the Weibull modulus is large enough,
$k \gtrsim \log V/V_0$. Clearly, the strength distribution flows away from the Weibull distribution in the limit of $V\to\infty$. 
The typical ranges for the Weibull modulus are $k > 30$ for metals, $5 < k < 20$ 
for ceramics~\cite{Tinschert2000529}, and $ 2 < k < 4 $ for biomaterials such as nacre~\cite{Menig20002383}. 
It is clear that for materials with small to moderate values of $k$ (such as biomaterials) the applicability of Weibull analysis is questionable. Indeed the weakest link hypothesis is manifestly
false --- these materials exhibit significant precursory fracture events (crackling noise) before
failure \citep{sethna2001}.
\par
\begin{figure}[tbp]
\begin{center}
{\includegraphics[width=0.4\textwidth,angle=0]{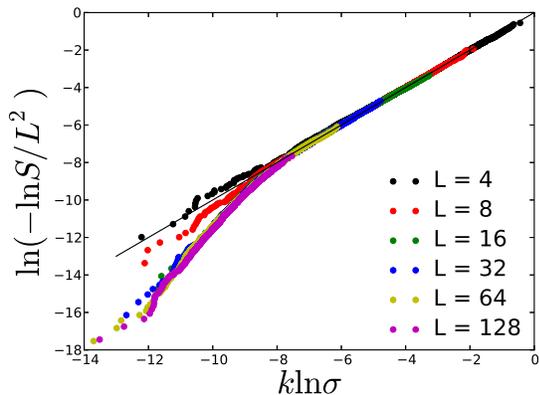}}
\end{center}
\caption{(color online) Emergent survival probability for $L=4,\ldots,128$, with $k=25$ for the threshold distribution. If the 
emergent distribution was Weibull, it would follow the solid black line. Clearly, the distribution flows away from Weibull 
at long length scales, showing that the Weibull distribution is not stable in a renormalization group sense.
}
\label{fig:WeibullFlow}
\end{figure}

Figure~\ref{fig:WeibullFlow} shows the emergent strength distribution for fuse networks of various sizes where  the fuse threshold is taken from a Weibull distribution with $k = 25$. 
We choose such a high value of $k$ to show the crossover away from Weibull, 
for smaller values of $k$ the distribution has already flown away from Weibull 
even for the smallest networks that we can simulate. 
According to the Weibull theory
the emergent distribution of strength would be given by Eq.~\ref{eq:SimWeibull} with $V/V_0 = L^2$ (there are $L^2$ fuses), thus 
giving $S_{L^2}(\sigma) = e^{-L^2\sigma^k}$. Figure~\ref{fig:WeibullFlow} shows that while this prediction holds 
for small values of $L$, the distribution flows away from the Weibull distribution at longer lengths. This shows 
that the Weibull distribution is unstable to disorder in a renormalization group sense, and must be used with 
caution for quasi-brittle materials.

We have established that the strength distribution flows away from Weibull in quasi-brittle materials,
but what does it flow towards? 
It is an unsolved problem to compute the new emergent distribution of strengths analytically.  
However, to get some idea about the distribution, we compute a very simple minded upper-bound to the survival probability
for the fuse network model. 
From Eq.~\ref{eq:StressIntensity}, at any given stress $\sigma$ 
the length of the critical crack goes as $w_{cr}(\sigma) \sim (\bar\sigma/\sigma)^2$ (i.e.~a crack 
longer than $w_{cr}$ will have unstable growth)
If the fuse strength threshold is standard Weibull, then the
probability of having a crack of size $w_{cr}$ at any given lattice site is at least 
$(1-e^{-\sigma^k})^{w_{cr}(\sigma)}$.
Since there are $L^2$ lattice sites, the global probability of survival is at most 
\begin{equation}
(1 - (1-e^{-\sigma^k}))^{w_{cr}(\sigma)})^{L^2}.
\end{equation}
Making asymptotic expansions for small $\sigma$, we get 
\begin{equation}
S_{L^2}(\sigma) < \exp(-L^2 e^{-k (\bar\sigma/\sigma)^2 \log (1/\sigma)} ).
\end{equation}
If we take the slowly varying $\log (1/\sigma)$ to be a constant, then the above expression is reminiscent DLB distribution\cite{DLB}.
The factor of $\log (1/\sigma)$ can be removed in a more natural way if one takes into account 
the stress concentration at each step of crack growth (see Ref.~\cite{Kahng1988} for a similar treatment). 
Our observation is supported by experimental results for some quasi-brittle materials where the DLB distribution was found 
to fit the data better than the Weibull distribution~\cite{hosson1991,alava1998,baxevanakis1993}.

\begin{figure}[tbp]
\begin{center}
{\includegraphics[width=0.4\textwidth,angle=0]{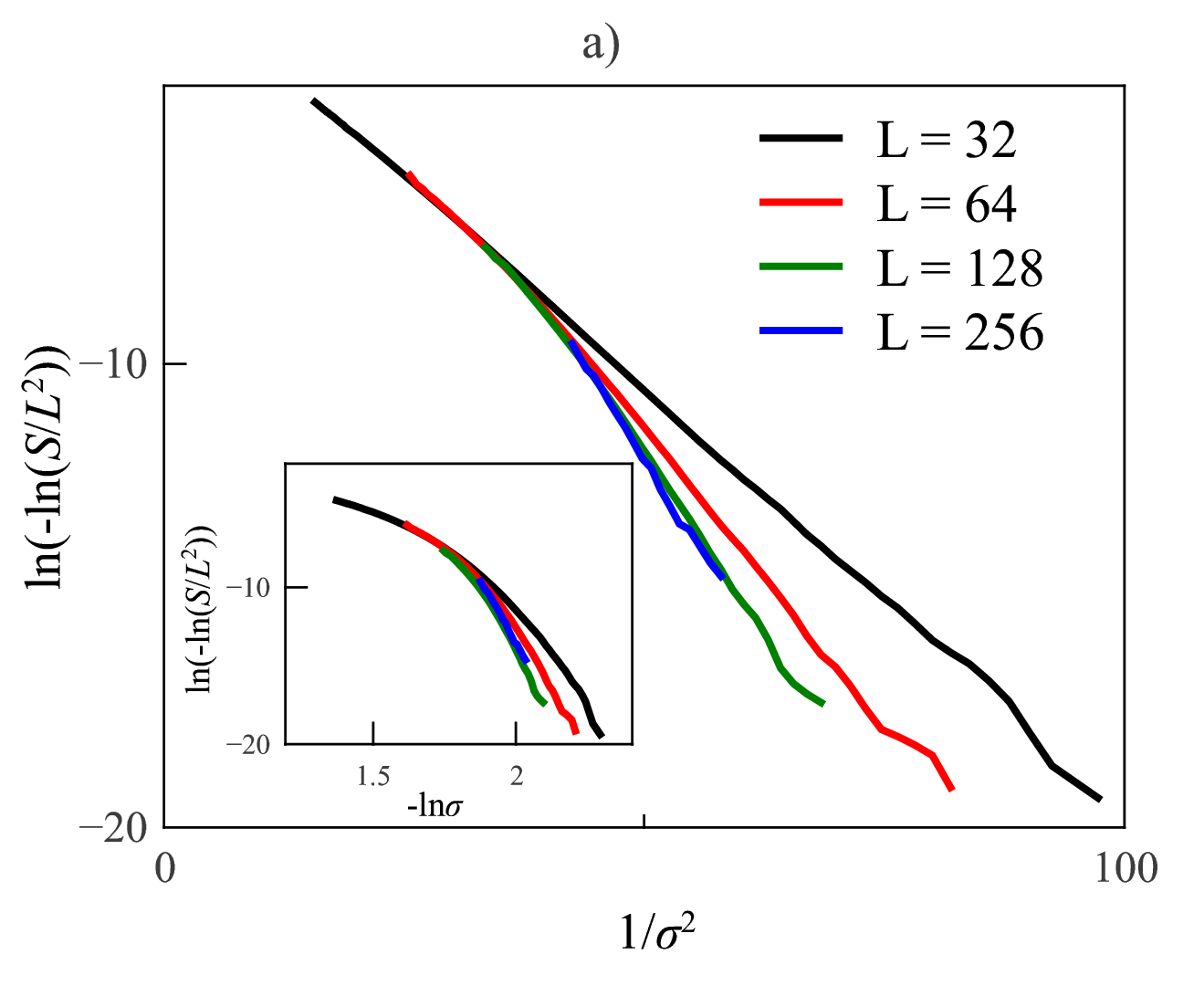}}
{\includegraphics[width=0.4\textwidth,angle=0]{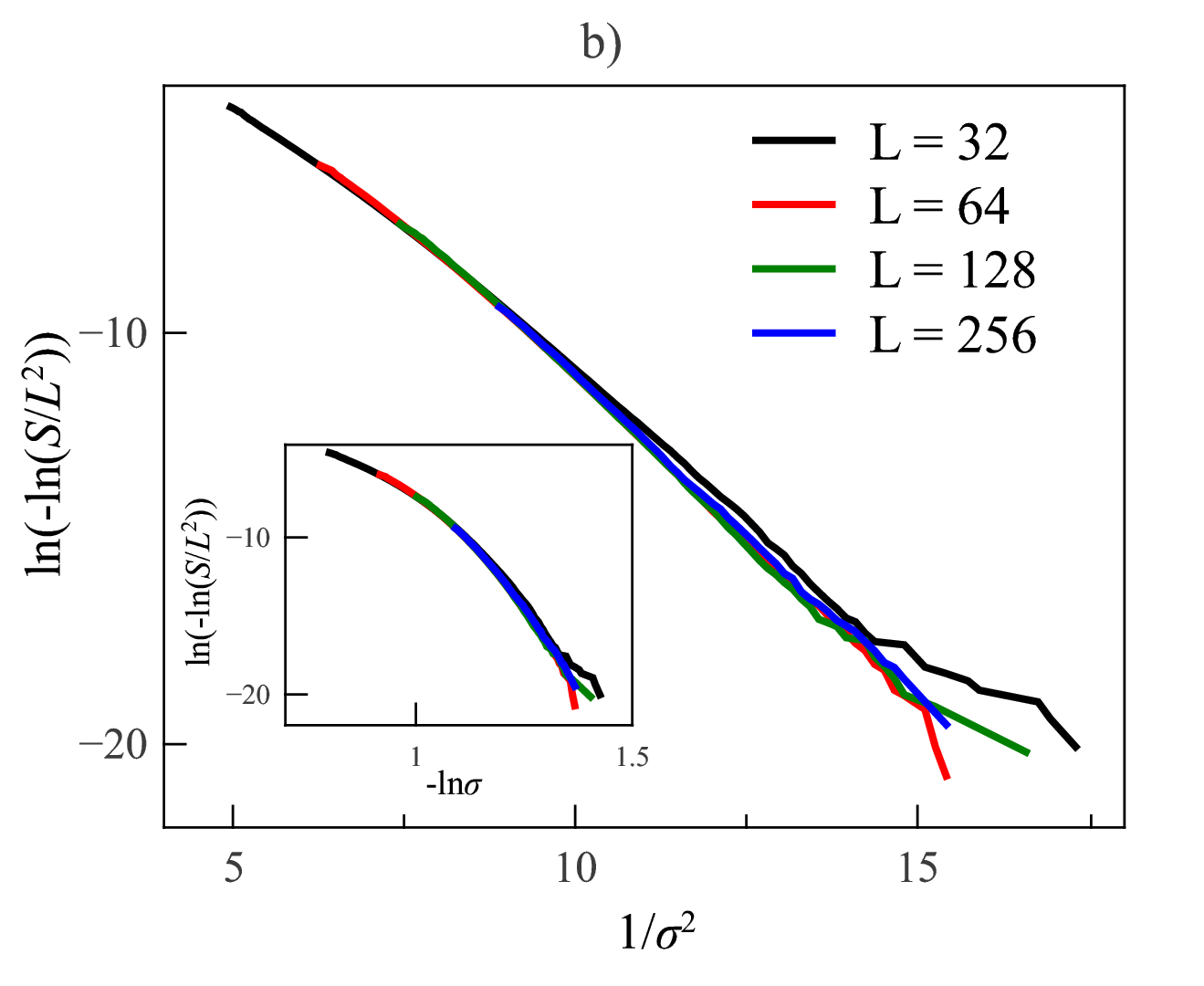}}
\end{center}
\caption{(color online) Survival probability for a) $k=1.5$ and b) $k=4$. The main figures show the DLB test, while the insets show the Weibull test; straight lines indicate agreement with the tested
form.}
\label{fig:Weibull}
\end{figure}

Since the upper bound that we have established decays faster than any Weibull function at $\sigma = 0$,  the macroscopic survival probability cannot be of the Weibull form, even 
if the microscopic distribution is Weibull. Note that the arguments made here are fairly 
general, and thus we expect the macroscopic strength distribution for any material with 
significant precursor damage to deviate from the Weibull distribution.
We have confirmed that these  ideas are consistent with the results of our numerical simulations. Figure~\ref{fig:Weibull} shows the survival probability  obtained by statistical sampling of fuse networks with different values of $k$. 
The main plot in the figure shows that the survival probability is consistent with a DLB distribution.
If instead the survival probability was consistent with a Weibull distribution, 
then the insets in the figure (so called Weibull plots) would be 
straight lines. However, the plots show considerable curvature, suggesting a deviation from the Weibull  distribution at long length scales.


\par

\begin{figure}[tbp]
\begin{center}
{\includegraphics[width=0.4\textwidth,angle=0]{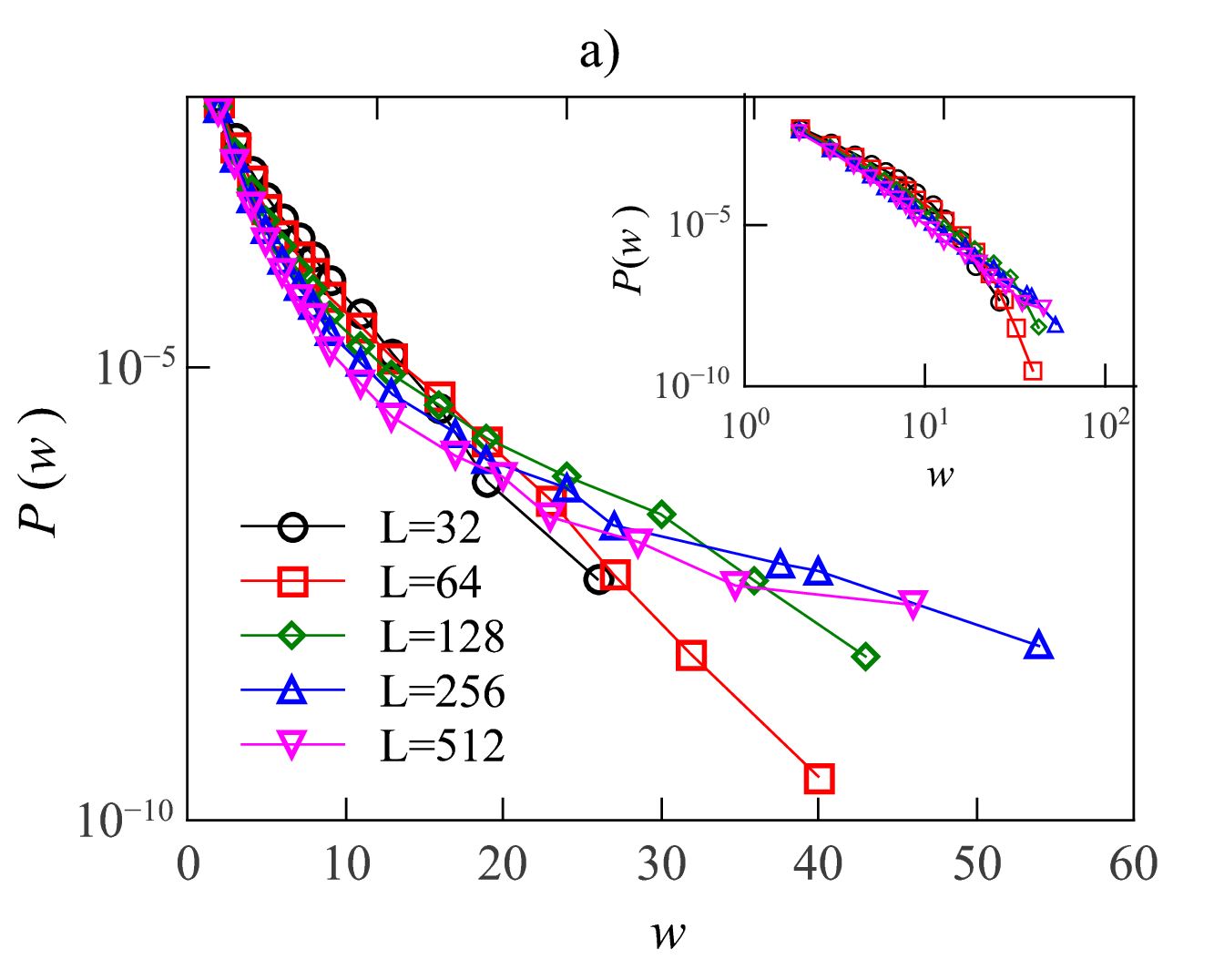}}
{\includegraphics[width=0.4\textwidth,angle=0]{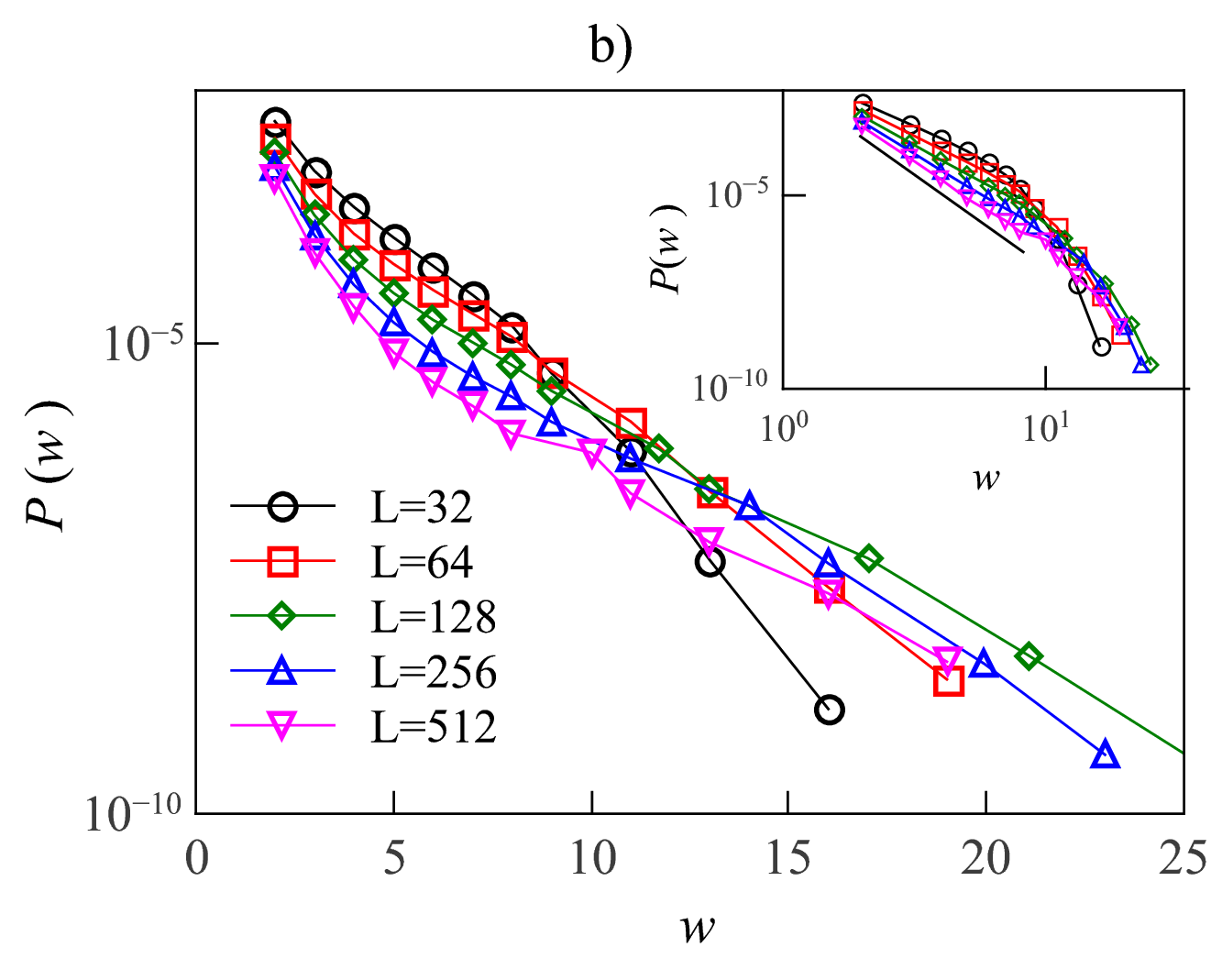}}
\end{center}
\caption{(color online) Crack-width distributions at peak load for Weibull-distributed fuse strengths with 
exponents a) $k=1.5$ and b)$k=4$. The distribution is a power-law with an exponential tail for all values of $k$. }
\label{fig:Weibull-crack}
\end{figure}

The DLB distribution was originally associated with samples having an exponential distribution
of crack length \citep{DLB,Manzato2012}. We have confirmed this hypothesis in 
our simulations by measuring the crack length distribution just before catastrophic failure.
The result, reported in Figure~\ref{fig:Weibull-crack} shows indeed the presence of an exponential tail.


\section{Defect distribution and Weibull modulus for Brittle materials}
\label{sec:Mod}
It is widely assumed that the emergent Weibull modulus for brittle materials
can be derived by using the Griffith's criteria if the crack length 
distribution is known. This assumption has been used in several 
important studies~\cite{Jayatilaka1977,danzer2007}. However, it has never been verified empirically due to 
experimental challenges. We examine this assumption numerically 
by simulating fuse networks seeded with power law distributed cracks. 
Cracks are created by removing a certain fraction, $p$, of fuses from the network.
The net density of cracks, $p$, is kept low ($< 0.2$) to mimic 
materials such as glasses or ceramics where the density of micro cracks is small.
The critical effects associated with approaching the percolation threshold are also 
avoided by keeping $p$ small. Unlike the classical fuse network models, 
the removed fuses are chosen so as to generate a power law distribution of crack lengths (section~\ref{sec:RFM}).
\par
We derive the strength distribution based on the standard Griffith's criteria based 
assumption and compare the result to simulations. 
According to Griffith's theory, 
if the exponent of the power law distribution of crack lengths is $\gamma$, 
then for Eq.~\ref{eq:DefectDensity} we have $h(w) \sim p w^{-\gamma}$, 
giving $f(\sigma) \sim p(\sigma/\bar\sigma)^{2\gamma}$,
where $\bar\sigma = K_{Ic}/Y$. This yields the following Weibull distribution 
of strengths for a fuse network of linear size $L$ and `volume' $L^2$ 
(assuming uniform stress)
\begin{equation}
S_{L^2}(\sigma) = e^{-L^2 p (\sigma/\bar\sigma)^{2\gamma}},
\end{equation}
thus the Weibull modulus is given by $k=2\gamma$.
\par
\begin{figure}[tbp]
\begin{center}
\includegraphics[width=0.4\textwidth,angle=0]{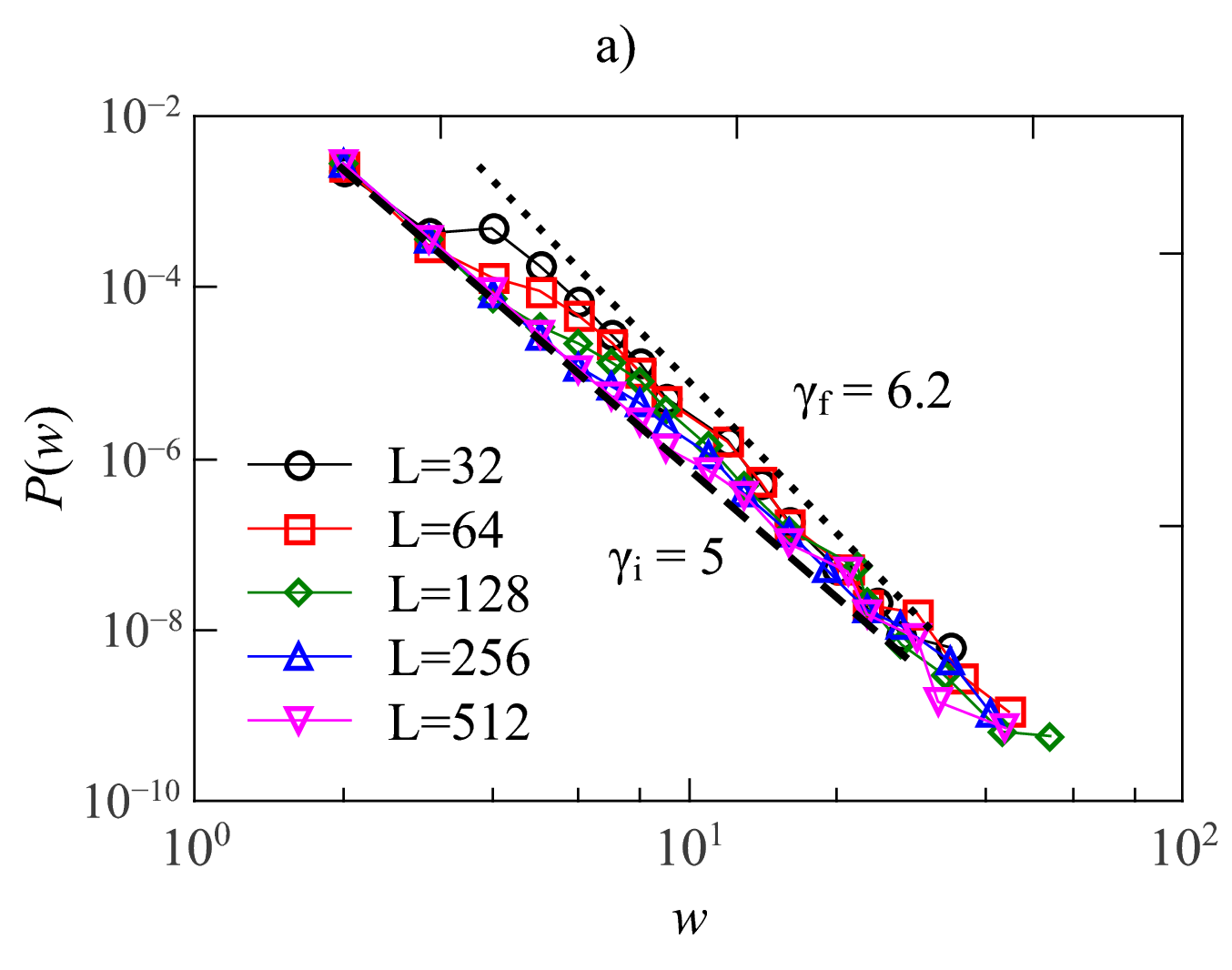}
\includegraphics[width=0.4\textwidth,angle=0]{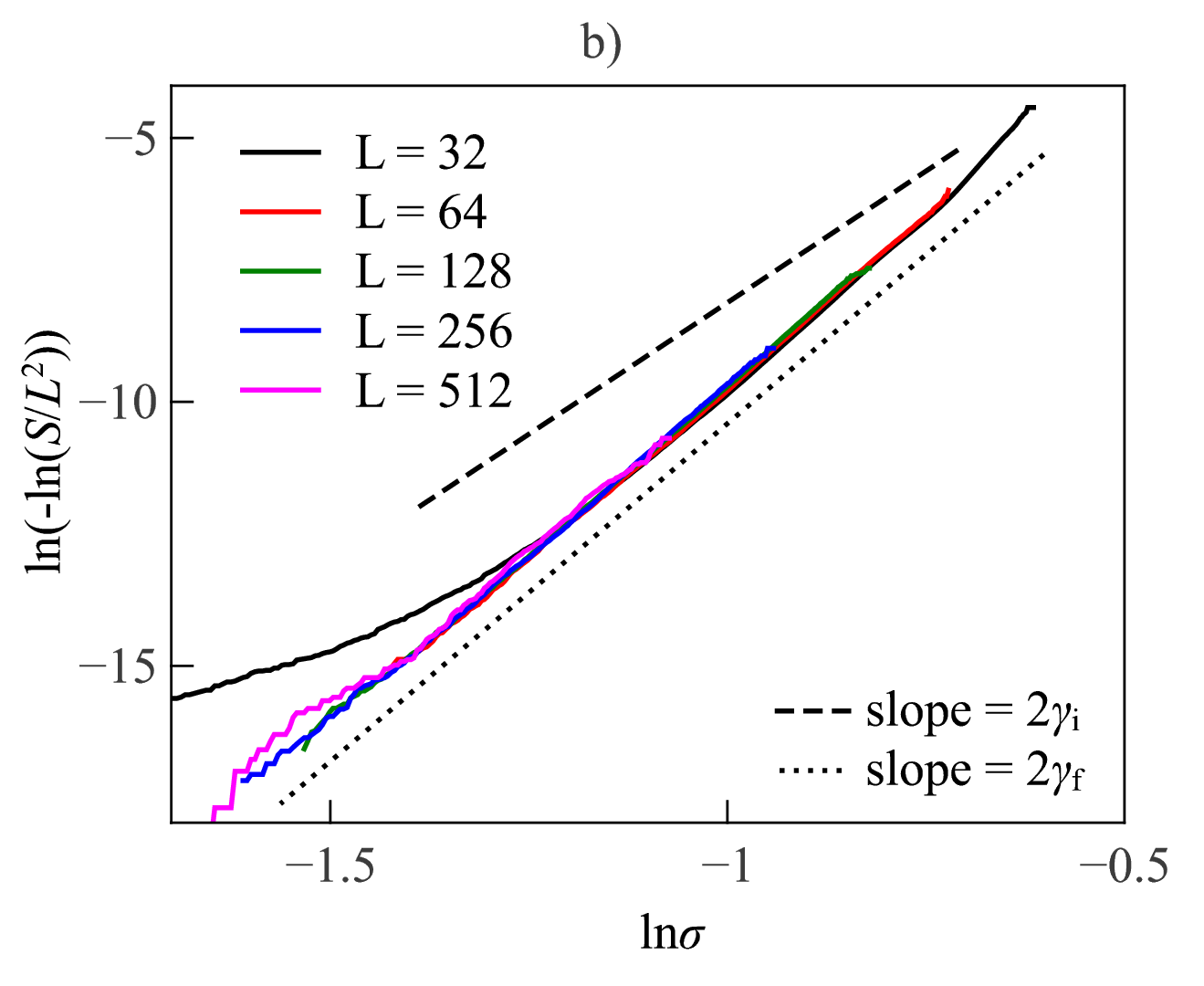}
\end{center}
\caption{(color online) a) Crack-width distributions at peak load for a system with power law distributed cracks. The power law tail
has an exponent $\gamma_{\rm f}$ that  is larger than the initial one $\gamma_{\rm i}$. b) The corresponding survival distribution obeys the Weibull law with $k=2\gamma_{\rm f}$}
\label{fig:powerlaw}
\end{figure}

The above discussion assumes that flaw distribution does not change at all 
in fracture process.
In real materials, as well as in our fuse network model, there is at least a small amount of 
damage before catastrophic fracture. 
This damage can change the tail of the crack width distribution. Let $\gamma_i$, $\gamma_f$ 
be the exponent of the crack size distribution before loading, and at peak load, respectively.
We investigate the relation between $\gamma_i$, $\gamma_f$, $p$, and $k$ numerically.
We find in our simulations that $\gamma_i < \gamma_f$. Further, we find that the modulus of the emergent Weibull distribution
is related to the damage distribution at peak load, $k = 2\gamma_f$. Figure~\ref{fig:powerlaw}a.~shows the comparison of the 
crack size distribution at zero and peak load for $\gamma_i = 5$. Figure~\ref{fig:powerlaw}b.~shows the corresponding survival probability 
on a so called Weibull plot. The slope of the Weibull plots agrees well with $2\gamma_f$. 


\par
\begin{figure}[ht]
\centering
\includegraphics[width=0.4\textwidth]{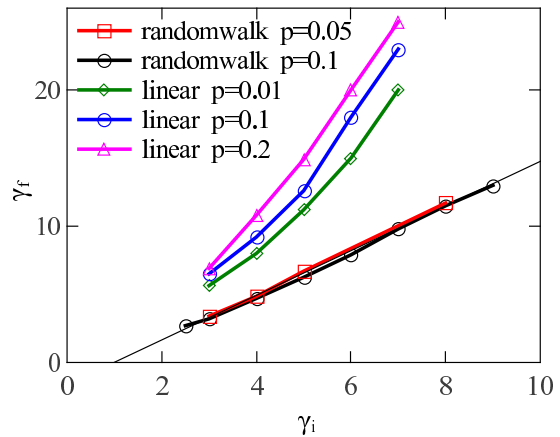} 
\caption{(color online) Relation of the exponents of the crack-width distribution initially, $\gamma_{rm i}$, 
and at peak load, $\gamma_{\rm f}$. For linear-grown cracks, the relation depends strongly on $p$ and $\gamma_{\rm i}$, 
while for random-walk-grown cracks we find $\gamma_{\rm f}\approx c \gamma_{\rm i} + d$, 
for both investigated dilution parameters $p=0.05$ and $p=0.1$. }
\label{fig:linexpos}
\end{figure}

Thus, the standard assumption of $k = 2\gamma_i$ is incorrect. 
We further explore the relation between $\gamma_i$ and $\gamma_f$ by carrying out extensive 
numerical simulations for $ 2.5 < \gamma_i < 9.0$, and $ 0.01 \leq p \leq 2$.
We also investigate the effect of the shape of initial cracks. We seed the network either with 
straight cracks, or fractal looking cracks grown by using self-avoiding random walks. In both cases 
we maintain the width distribution, $h(w) \sim w^{-\gamma_i}$, and the defect density as dictated by
$p$. Figure~\ref{fig:linexpos} shows the relation between $\gamma_i$ and $\gamma_f$ for various values
of $p$ for straight as well as grown cracks. For all the cases we observe that $\gamma_f > \gamma_i$. 
It is reasonable to expect a slight increase in the 
exponent $\gamma$ due to crack bridging. However, it is not clear what causes the almost three times
increase in $\gamma$ for some configurations. 

\section{T-method to fit the strength distribution}
\label{sec:TMethod}
Except for the case of power law distributed cracks, we see that the 
strength distribution is not Weibull, and is probably of the DLB type. 
In a previous paper~\cite{Manzato2012} 
we discussed how the extreme value functions are 
an extremely poor approximation to the DLB form.
These considerations raise the following question: what form should be used to fit fracture data in practice?
\par
\begin{figure}[htbp]
\centering
\includegraphics[width=0.4\textwidth]{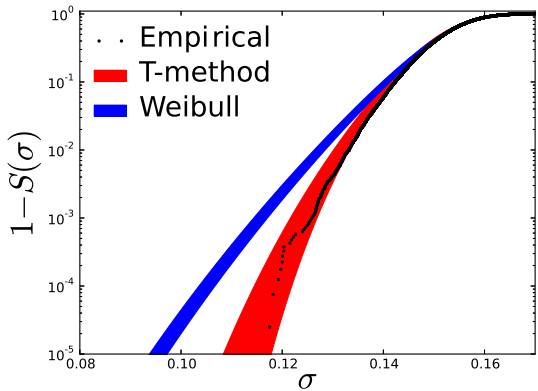}
\caption{(color online) Results of extrapolating the fits of the Weibull distribution and the suggested 
transformation based method. Fits obtained from small datasets (size 200) by using the T-method
can be extrapolated with confidence to probabilities about 2 orders of magnitude smaller. The figure
shows $\pm$1 standard deviation results.}
\label{fig:transformation}
\end{figure}

One of the major concerns while fitting data to extreme value 
distributions is the accuracy of extrapolations in the low probability tail. 
We compare the standard Weibull theory and the recently proposed T-method~\cite{shekhawat2014} 
by fitting the data fracture data for the quasi-brittle fuse networks with the two techniques.
In Weibull theory (Eq.~\ref{eq:Weibull}) the survival probability of the network 
is given by $S_{L^2}(\sigma) = e^{-L^2(\sigma/\bar\sigma)^k}$. Given the observed 
data vector $\mathbf{X}$ (= vector of fracture strengths observed in simulation) of length $n$, the parameters
$(\bar\sigma, k)$ are determined by using the maximum likelihood estimation (MLE) as the values
that maximize the following log-likelihood function
\begin{equation}
\mathcal{L}_W(\bar\sigma,k|\mathbf{X}) = \sum_{i=1}^{n} \partial_{\mathbf{X}_i}\log S_{L^2}(\mathbf{X}_i).
\label{eq:WeibullMLE}
\end{equation}
The parameters that minimize the above log-likelihood function give the best fit 
parameters $(\bar\sigma,k)$ for the Weibull theory. The T-method first applies 
a nonlinear transformation to the data, $T(\mathbf{X}) = \mathbf{X}^{-\alpha}$, 
and then fits the transformed data of an extreme value form, thus giving 
the following log-likelihood function~\cite{shekhawat2014}
\begin{equation}
\mathcal{L}_T(\alpha, a, b|\mathbf{X}) = \sum_{i=1}^{n} \log \partial_{\mathbf{X}_i}G_0((T(\mathbf{X}_i)-b)/a),
\label{eq:TMLE}
\end{equation}
where the parameters $(\alpha,\ a,\ b)$ are estimated by minimization, and $G_0(x) = \exp(-e^{-x})$ is the standard Gumbel distribution.
We use the dataset 
of over 20,000 simulations, corresponding to the random fuse model with $k=1.5$ and $L=128$ 
to test the applicability of the above method for such extrapolations.
We choose 20 random samples of 200 data points from the dataset. We then fit each of the 
smaller data sets using the Weibull theory and the T-method. We extrapolate the fits and compare prediction 
in the low probability tail with the empirical data. Figure~\ref{fig:transformation} shows the $\pm$1 standard 
deviation predictions of such fits. It is clear from the figure that the T-method outperforms the 
standard Weibull theory in accuracy of the fit and it extrapolation in the low stress tail.

\section{Conclusions}
\label{sec:conc}
In conclusion, we have studied the conditions for emergence of the Weibull distribution for fracture strength in brittle and quasi-brittle materials. 
We show the Weibull distribution is unstable under coarse graining for a large class of materials where the 
weakest link hypothesis is not strictly valid, and there is significant precursor damage. 
For the case of brittle materials we show that the relation between strength distribution 
and the defect size distribution is highly non-trivial and cannot be obtained by simple application of the Griffith's criteria. 
Crack bridging has significant effect on the tails of the crack size distribution, and thus changes the Weibull modulus considerably. 
We find that the recently proposed T-method does a significantly better job at fitting the fracture strength data, as compared 
to the Weibull distribution.
We hope that the our results will lead to further research and discussion
about the applicability of the Weibull distribution for fracture data, particularly for quasi-brittle materials that crackle.

We thank Zoe Budrikis and Claudio Manzato for valuable comments and fruitful discussions. 
Z. B. and S. Z. were supported by the  European Research Council through the Advanced Grant 2011 SIZEFFECTS. A.S.~was partially supported by the Miller Institute for Basic Research in Science, Berkeley. J.P.S.~and A.S.~(partially) were supported by DOE-BES DE-FG02-07ER46393. J.P.S. and S.Z acknowledge support from Materials World Network: Cooperative Activity in Materials Research between US Investigators and their Counterparts Abroad in Italy (NSF DMR 1312160).

\end{document}